\documentstyle[definitions,sprocl]{article}

\def\be{\begin{equation}}
\def\ee{\end{equation}}
\def\bea{\begin{eqnarray}}
\def\eea{\end{eqnarray}}

\def\di{\displaystyle}

\def\bg{\begin{eqnarray}\begin{array}{rcl}\displaystyle}
\def\eg{\end{array} &\di    &\di   \end{eqnarray}}
\def\bm#1{\begin{eqnarray}\begin{array}{#1}\di}
\def\bgo{\begin{eqnarray*}\begin{array}{rcl}\displaystyle}
\def\ego{\end{array} &\di    &\di \nonumber  \end{eqnarray*}}

\def\btensor#1#2{\renew\left#1\begin{array}{#2}\di}
\def\brtensor#1#2#3{\ren#3\left#1\begin{array}{#2}}
\def\botensor#1#2{\renew\left#1\begin{array}{#2}}
\def\etensor#1{\end{array}\right#1}

\def\eq#1{(\ref{#1})}
\def\Eq#1{Eq.~(\ref{#1})}

\def\d{{d}}

\def\Tr{\mbox{Tr}}

\def\ov{\over}

\def\del{{\mbox{\boldmath$\delta$}}}

\def\R{\mbox{l}\!\mbox{R}}

\def\CN{{\cal N}}

\def\CW{{\cal W}}

\date{\today}

\def\ren#1{\renewcommand{\arraystretch}{#1}}
\def\rene{\renewcommand{\arraystretch}{1.9}}
\def\renew{\renewcommand{\arraystretch}{1}}
\rene

\begin{document}
\thispagestyle{empty}
${}$\vskip 8em

\begin{center}
{{\bf On Gauge Invariant Wilsonian Flows}\footnote{Based on seminars 
presented at the workshop on the Exact Renormalisation Group, Faro, 
September 1998. To be published in the proceedings.} \par}       
   \vskip 3em                      
{\rm Daniel F. Litim}\par
\vskip .5em 
{\it Departament ECM {\sl \&} IFAE,
Facultat de F\'{\i}sica, Universitat de Barcelona\\ Diagonal 647, E-08028 
Barcelona, Spain. E-Mail: Litim@ecm.ub.es}\par
\vskip 1.5em 
{\rm Jan M. Pawlowski}\par
\vskip .5em 
{\it Dublin Institute for Advanced Studies,
10 Burlington Road\\ Dublin 4, Ireland. E-Mail: jmp@stp.dias.ie}
\end{center} \par
\vskip 3em

\begin{center}
{\footnotesize\bf  Abstract}\\ \vskip .5em
{\begin{minipage}{4.2truein}
\parindent=0pt 
\footnotesize
{We investigate non-Abelian gauge theories within a 
Wilsonian Renormalisation Group approach. Our main question is: How close 
can one get to a gauge invariant flow, despite the fact that a Wilsonian 
coarse-graining seems to be incompatible with gauge invariance? We discuss 
the possible options in the case of {\it quantum} fluctuations, and argue 
that for {\it thermal} fluctuations a {\it fully} gauge invariant 
implementation can be obtained.}\par
                 \end{minipage}}
\end{center}

\vspace*{-12cm}
\begin{flushright}
{\normalsize DIAS-STP-99-01\\ ECM-UB-PF-99-02 }
\end{flushright}
\vspace*{10cm}

\newpage
\setcounter{page}{1}
\section{Introduction}
The Wilsonian Renormalisation Group \cite{Wilson} 
has proven itself to be a powerful tool for studying both 
perturbative and non-perturbative effects in quantum field theory. It has 
been particularly successful for scalar theories, where a number of new 
results have been obtained.
One expects that a suitable formulation for non-Abelian 
gauge theories might provide new insight into non-perturbative 
effects in QCD as well. 
However, the Wilsonian approach is based on the concept of a step-by-step 
integrating-out of momentum degrees of freedom and one may wonder whether this 
concept can be adopted for gauge theories. Indeed, one of the main obstacles 
of the Wilsonian approach to gauge theories was precisely the question as 
to how  gauge invariance can be controlled.

In this contribution we analyse this question in some detail. In 
particular, we want to understand how close one might come to a gauge 
invariant implementation of a Wilsonian coarse-graining in the usual sense and 
which net gains are related to such procedures. 

This contribution is organised as follows: The first parts consider generic 
features of Wilsonian flows for gauge theories. We discuss 
the modified Ward Identities, in particular the resulting constraints
 for a construction of gauge invariant flows, and we detail the link 
between the standard approach and the background field methods. Finally 
we comment on the key features of the different gauge invariant approaches. 
In the 
remaining part, we investigate the application to thermal field theory, 
and argue why a gauge invariant implementation is possible for all scales 
in the case of thermal fluctuations.

\section{The flow equation}
We investigate these questions in a path integral 
approach based on ideas of Polchinski \cite{ERG}. In this approach 
a momentum cut-off is achieved by adding a cut-off term $\Delta_k S$ 
to the action which 
is quadratic in the field. This results in
 an effective action $\Gamma_{k}$ where momenta larger 
than $k$ have been 
integrated-out. The change of $\Gamma_k$ under 
an infinitesimal variation of the scale $k$ is described by a flow equation
which can be used to successively integrate-out the momenta
smaller than the cut-off scale $k$. Thus given an effective action 
$\Gamma_{k_0}$ at an initial scale $k_0$ the flow equation provides us with a 
recipe how to calculate the full effective action $\Gamma$.  

The introduction of $\Delta_k S$ seems to 
break gauge invariance. However, $\Gamma_k$ satisfies a modified
Ward identity (mWI). This mWI commutes with the flow and 
approaches the usual Ward identity (WI) as $k\to 0$. 
Consequently the full effective action $\Gamma$ satisfies the usual Ward 
identity. In other words, gauge invariance of the full theory is preserved 
if the effective action $\Gamma_{k_0}$ satisfies the mWI 
at the initial scale $k_0$. Let us see how close to a gauge invariant 
description one might get even during the flow, that is for all $k$. 
\subsection{Derivation of the flow equation}\label{local}
To that end let us outline the derivation of the flow equation.  
In the following we employ the superfield formalism as introduced in 
\cite{heavy}. A more explicit version of the following arguments 
may be found in \cite{Axial,analytic2,reutwett,collect}. 
The starting point is the infrared (IR) regularised Schwinger functional 
\bm{c}\label{schwingerk}
\exp W_k[J]={1\ov \CN_k}\int \d \phi\,\exp\{-S_k[\phi]+
\Tr\, \phi^* J\}, 
\eg 
where the trace $\Tr$ denotes 
a sum over momenta, indices and the different fields $\phi$, including a minus 
sign for fermionic degrees of freedom. $(\phi_i)=(\phi_1,...,\phi_s)$ 
is a short-hand notation for all fields and $\phi^*$ denotes its dual. 
The term $S_k[\phi]$ contains the 
(gauge-fixed) classical action and a quadratic cut-off term 
$\Delta_k S[\phi]$ given by (e.g.\ \cite{Axial} and 
references therein): 
\bg\label{cut-off1} 
\Delta_k S[\phi] = \frac{1}{2} \Tr\, \left\{
\phi^* R^{\phi}_k[P_\phi]\phi\right\},
\eg 
where $P_\phi^{-1}$ is proportional to 
the bare propagator of $\phi$. The $k$-dependent constant $\CN_k$ has been 
introduced to guarantee an appropriate normalisation of $W_k$. 
The flow of \eq{schwingerk} related to an infinitesimal change of $k$ 
is given by ($t=\ln k$) 
\bm{c}\label{floww}
\partial_t \exp W_k[J]= -{1\ov \CN_k}\int \d \phi\left( {1\ov 2}
\Tr\, \phi^* R_k^\phi\phi
+{\partial_t \CN_k\ov \CN_k}\right) \,e^{-S_k[\phi]+
{\rm Tr}\, \phi^* J} . 
\eg  
Eventually we are interested in the Legendre transform of $W_k$, 
the effective action $\Gamma_k$:  
\bm{c}\label{Gk}
\Gamma_k= \Tr\, \phi^* J-W_k[J]-\Delta_k S[\phi]. 
\eg 
For the effective action $\Gamma_k$ the flow equation \eq{floww} turns into  
\bm{c}\label{flow}
\partial_t\Gamma_k[\phi]=  \frac{1}{2}\Tr\left\{ G^{\phi^*\phi}_k[\phi]
\partial_t R^\phi_k[P_\phi]\right\}+\partial_t \ln \CN_k 
\eg 
with 
\bm{c}\label{props}
G^{\phi^*_i \phi_j}_k[\phi]=  
\left(\frac{\delta^2\Gamma_k[\phi]}{\delta\phi^*_i \delta\phi_j} + 
R_{k,ij}^{\phi}[P_\phi]\right)^{-1}. 
\eg 
The flow equation 
\eq{flow} and an initial effective action $\Gamma_{k_0}$ 
may be read as a proper definition of the path integral.  
In order to have a well-defined flow equation we use  
operators $R_k^\phi$ with the following properties 
($x=P_\phi\, k^{2 d_\phi-d}$): 
\begin{eqnarray} \label{limits}
k^{2 d_\phi-d} 
R_k^{\phi}[P_\phi] &\di \stackrel{x\rightarrow 0}{\longrightarrow} 
&\di \frac{x^n}{|x|}, \ \ \ \  n\leq 1\label{property1}
\\\di 
x^{d/(2 d_\phi-d)} P_\phi^{-1} R_k^\phi[P_\phi]
&\di \stackrel{{x}\rightarrow \infty}{\longrightarrow}&\di   0,
\label{property2}
\end {eqnarray}
where $d_\phi$ are the dimensions of the fields $\phi$. A very 
interesting subset of operators $R_k$ with the properties 
\eq{property1},\eq{property2} is the set of operators which 
satisfy $n=1$ in \eq{property1} and decay 
exponentially for $x\to \infty$. 
These lead to very good convergency 
properties if invoked for numerical calculations. 
A cut-off term \eq{cut-off1} with $R_k$ satisfying 
\eq{property1},\eq{property2}
effectively suppresses modes with momenta $p^2\ll k^2$ in the 
generating functional. 
For modes with large momenta $p^2\gg k^2$ the cut-off term 
vanishes and in this regime the theory remains unchanged. In the limit 
$k\rightarrow 0$ we approach the full generating functional $\Gamma$ since 
the cut-off term is removed. In the limit $k\to \infty$ all momenta are 
suppressed and --with a suitably chosen $\CN_k$-- the effective action 
approaches the (gauge fixed) 
classical action $S_{\rm cl}+S_{\rm gf}$. 
Hence $\Gamma_k$ interpolates between the 
classical action and the full effective action: 
\bg\label{0infty}
S_{\rm cl}+S_{\rm gf}\stackrel{k\to\infty}{\longleftarrow}
\Gamma_k \stackrel{k\to 0}{\longrightarrow} \Gamma.
\eg 
Note that $\partial_t R_k^\phi$ 
serves as a smeared-out $\delta$-function in momentum space peaked at about 
$p^2\approx k^2$. In this sense, a Wilsonian flow is local, as only a 
small window of momentum modes around $k$ will contribute to the flow. 
By varying the scale $k$ towards smaller $k$ 
according to \eq{flow} one successively integrates-out 
momentum degrees of freedom. 

\section{Control of gauge invariance}\label{gaugeInv}
In this section we show how gauge invariance is --in general-- controlled 
by modified Ward Identities (mWI). For perturbative 
truncations to the effective action $\Gamma_k$, the mWI can 
be maintained employing the quantum 
action principle. The more important point is that the mWI can also 
be satisfied within non-perturbative truncations. 

\subsection{Modified Ward Identities}\label{control}
When applied to gauge theories the cut-off term \eq{cut-off1} generates 
additional terms in the Ward identity thus leading to a modified Ward 
Identity. For non-Abelian gauge theories formulated in general linear gauges 
$L_\mu A_\mu$ and gauge fixing parameter $\xi$, 
one can derive the following mWI \cite{Axial,analytic1} 
\bm{rl}\label{mwi}
\CW_k[\phi]=&\di  
\del_\alpha 
\Gamma_{k}[\phi]-
\Tr\,(L_\mu D_\mu \alpha)\,\frac{1}{\xi}\,L_\nu A_\nu 
+\frac{g}{2}\Tr\,\left[\alpha,\left( 
R_k^{\phi}+L^*\frac{1}{\xi}\,L\right)\right] G^{\phi^*\phi}_{k}\\\di 
= &\di 0,  
\eg
where we used the abbreviation $\del_\alpha$ for the generator of 
infinitesimal gauge transformation on $\phi$, e.g.\ 
$\del_\alpha A= D(A)\alpha$ for the gauge field and $\del_\alpha \psi=
\alpha\psi,\ \del_\alpha \varphi=\alpha \varphi$ for fermions and scalars 
respectively. Note also that $\alpha$ has to be 
taken in the representation of the corresponding field $\phi_i$. For the 
sake of brevity we dropped any reference to possible ghost fields which have 
to be added in general; however they have to be treated in a similar manner. 

The cut-off dependent terms in \eq{mwi} vanish for $k\to 0$. The 
compatibility of the mWI \eq{mwi} with the flow equation is given by
\cite{Axial,analytic1}  
\bm{c}
\partial_t \CW_k[\phi]=
-\frac{1}{2}\Tr\left( G^{\phi^*_i\phi_j}_k \partial_t 
R^{\phi_j}_k G^{\phi^*_j\phi_l}_k \frac{\delta}{\delta \phi^*_l}
\frac{\delta}{\delta \phi_i}\right)\CW_k[\phi]. 
\label{compatible}\eg 
\Eq{compatible} ensures that if the initial 
effective action satisfies the mWI \eq{mwi} than the usual Ward identity is 
satisfied for $k=0$. It is now possible to employ the 
quantum action principle. One can solve the mWI \eq{mwi} 
order by order in the coupling. This solution will remain a solution for
any $k$ due to \eq{compatible}. 
In other words, an effective action $\Gamma_{k_0}$ that solves the mWI 
up to order $n$ in the coupling stays a solution up to order $n$ for any 
$k_0$, in particular for $k=0$. This is how perturbation theory is 
included as one possible truncation scheme (see also \cite{datmorris}).  

In general one may use another expansion parameter --instead of 
the coupling-- which is small in the regime of interest, e.g.\ 
$p^2/\Lambda^2_{QCD}$ in the deep infrared regime of QCD. 

\subsection{Numerical implementation}\label{numerics}
The flow equation \eq{flow}, the mWI \eq{mwi} and a suitably truncated
 (initial) effective action are the starting points for 
numerical applications \cite{heavy,berwett}.  
The first step is to introduce a parametrisation of the effective action 
in terms of some couplings $\gamma$. In an expansion in the powers of the 
fields, these couplings are just the momentum-dependent vertex functions. 
The key point is that the mWI introduces relations between 
the different couplings $\gamma$. Thus only a subset of couplings 
$\{\gamma_{\rm ind}\}$ can be independently fixed, whereas the other couplings 
$\gamma_{\rm dep}$ can be derived with the mWI and the set 
$\{\gamma_{\rm ind}\}$. It is worth mentioning that 
the splitting $\{\gamma\}=\{ \gamma_{\rm ind},\gamma_{\rm dep}\}$ 
is not unique. 

Now we chose a truncation of the flow equation such that only the 
couplings $\gamma$ related to operators  
important for the problem under investigation are included. 
For QCD, these typically include the gauge coupling, a gluonic mass 
term, and the 3- and 4- point (and higher) vertices. 
In general, only a 
(finite) subset of $\{\gamma_{\rm dep}(k)\}$ is taken into account. 
The flow equation will then be integrated as follows: 
After integration of an infinitesimal momentum shell between $k+\Delta k$ and 
$k$ we obtain couplings $\gamma_{\rm ind}(k)$. With the mWI one derives the 
(finite) subset of $\{\gamma_{\rm dep}(k)\}$ which 
together with the $\gamma_{\rm ind}(k)$ serve as the input for the next 
successive integration step. By employing this procedure  
for the integration of the (truncated) flow equation from the 
initial scale $k_0$ to a scale $k$ one obtains a set of 
$\{\gamma(k)\}=\{\gamma_{\rm ind}(k),\gamma_{\rm dep}(k)\}
$ for any scale $k$. These $\gamma(k)$ parametrise an effective action 
$\Gamma_k$ which by construction does satisfy the mWI at any scale $k$, 
in particular for $k=0$ (see section~\ref{control}). Thus the full 
effective action $\Gamma_{\rm trunc}$ 
calculated with the truncated flow equation satisfies the 
usual Ward identity. The truncation does not imply a breaking of gauge 
invariance 
but rather a neglecting of the back-reaction of the truncated couplings on 
the flow of the system. 

For a validity check of the truncation we have to employ the fact that the 
system is overdetermined. The set $\{\gamma_{\rm dep}\}$ 
may be also directly calculated
 with the flow equation \eq{flow} itself. Only for the full 
system both equations (flow equation and mWI) are compatible, 
as shown in section~\ref{control}, \Eq{compatible}. 
Thus as long as the results for the $\gamma_{\rm dep}$, 
which are obtained by either using \eq{flow} or using 
\eq{mwi}, do not 
deviate from each other, the truncation remains valid. The validity bound of
a truncation is reached when these independently determined results for 
$\gamma_{\rm dep}$ no longer match. Typically, this defines a final
cut-off scale $k_{\rm fin}\ll k_0$. Such a check has 
been done with the gluonic mass \cite{heavy}. 
Even though this was only a partial consistency check --and 
thus not entirely satisfactory-- it essentially gives the flavour 
of what has to be done in practice: For 
non-perturbative truncations the mWI is employed both for the 
consistency check {\it and} as a tool in order to calculate the value of the 
$\gamma_{\rm dep}$. 
As mentioned in the last section, 
for a fully controlled calculation one additionally has to find a suitable 
expansion parameter which can be employed for general validity checks of 
the truncation. 

\section{Analytic methods}
In this section we detail how the flow equation can be employed to derive 
non-trivial analytic results. We make use of the 
background field approach, and, subsequently, of general axial gauges. 

\subsection{Background field formalism}\label{appl}
Let us 
first show how the usual Ward identity is obtained by using a background 
field \cite{analytic1}. The following derivation is strictly valid only for
 momentum independent gauges; however a minor modification of it also applies 
to general linear gauges \cite{analytic2}. 
The background field dependence in this approach 
is given by an equation quite similar to the flow equation: 
\bm{c}
\frac{\delta }{\delta \bar A}\partial_t\Gamma_k=
\frac{1}{2}\partial_t
\Tr \left\{G^{\phi^*\phi}_k[\phi,\bar A]{\delta\ov \delta 
\bar A} R^{\phi}_k\right\}. 
\label{bar A}\eg 
This translates into the following equation for an infinitesimal 
gauge transformation of $\bar A$ applied on the effective action: 
\bm{c}\label{barwi}
\bar\del_\alpha\Gamma_k[\phi,\bar A]= \frac{g}{2}\Tr\left\{ 
\,[\alpha, R_k^{\phi}] G^{\phi^*\phi}_{k}\right\}, 
\eg 
where $\bar\del_\alpha$ is defined by its action on the fields, e.g.\  
$\bar\del_\alpha =D(\bar A)\alpha,\ \bar\del_\alpha\phi=0$. It follows from 
\eq{mwi}, \eq{barwi} that 
$\Gamma_k[\phi,\bar A]-S_{\rm gf}[A]$ is invariant under the transformation 
$(\del_\alpha+\bar\del_\alpha)$. As a consequence 
$\hat\Gamma_k[\phi]:=\Gamma_k[\phi,\bar A=A]$ satisfies the 
usual WI without the cut-off dependent terms: 
\bm{c}\label{wi}
\del_\alpha \hat\Gamma_k[\phi] =
\Tr\ [L_\mu D_\mu(A)\alpha]\, \frac{1}{\xi} L_\nu A_\nu. 
\eg 
where the gauge field derivative involved in \eq{wi} hits both the 
gauge field $A$ and the auxiliary field $\bar A=A$. Note however that 
the propagator $G^{AA}_k$ is still the one derived from 
${\delta^2\ov \delta A^2}\Gamma_k[\phi,\bar A]$ at $\bar A=A$. The flow 
equation for $\hat\Gamma_k$ requires the knowledge of $G^{AA}_k$, thus 
slightly spoiling the advantage of dealing with an effective action 
which satisfies the usual WI even for $k\neq 0$. 

\subsection{General axial gauges}

General axial gauges have recently been studied within this approach
\cite{Axial,analytic2,analytic1}. It has been established that 
the spurious 
singularities of perturbation theory are absent \cite{Axial}. This implies that
the theory is well-defined without any further regularisation in contrast to 
standard perturbation theory. 
Other advantages are the absence of Gribov copies and the decoupling of the 
ghost sector. 
Finally, the gauge fixing parameter $\xi$ has a non-perturbative fixed 
point at $\xi=0$, which makes this formulation very attractive for both 
analytical and numerical computations. 

In the following we refer to analytic calculations done in a general 
axial gauge and using the background field approach as 
briefly discussed in section~\ref{appl}. As a consequence of \eq{wi} 
we have gained gauge invariance --apart from the explicit breaking from 
the gauge fixing-- even for $k\neq 0$. This 
simplifies the expansion of the effective action. The 
problem is now to distinguish between the gauge field $A$ and the 
field $\bar A=A$ which only serves as an auxiliary variable. This is 
necessary since the flow equation still requires the knowledge of $G_k^{AA}$ 
as mentioned above, and which can be obtained at least in principle from 
\eq{bar A}. Additionally the $\bar A$-dependent of the effective action 
should be dropped completely since in the present approach it is {\it only} 
an auxiliary variable and for $k=0$ there is no $\bar A$-dependence at all.  

Let us illustrate the importance of the latter point  
with the following example of the 1-loop $\beta$-function: 
It can be shown with \eq{bar A} (apart from higher order 
terms in $\bar A$) that the flow equation on 1-loop level 
leads to a contribution proportional to $(n-1)\int F^2(\bar A)$ 
to $\Gamma_k$ \cite{analytic2}, where $n$ has been defined in 
\eq{property1}. 
This vanishes for operators $R_k$ with masslike IR limit ($n=1$) even though 
it is in general non-zero beyond 1-loop level. For these $R_k$ the 
1-loop $\beta$-function is straightforwardly calculated. 
However for $R_k$ satisfying \eq{property1} 
with $n\neq 1$ this term is non-vanishing and
 has to be subtracted from $\hat\Gamma_k$ as calculated 
with the flow equation. In principle this can be done by using an 
appropriately defined $\CN_k$. 
Subtracting the contribution proportional to 
$(n-1)\int F^2(\bar A)$ from the flow equation 
leads to the correct 1-loop $\beta$-function and other 1-loop quantities. 

With \eq{flow}, \eq{mwi}, \eq{bar A} and \eq{barwi} we can 
investigate the effective action analytically. It is worth noting that the 
flow equation is a `1-loop' equation, even though the loops depend on the full 
field dependent propagator. Thus heat kernel methods can be employed, 
although {\it not} as a regularisation method, since everything is finite 
from the onset. Restricted to the perturbative regime, these calculations 
\cite{Axial,analytic2,analytic1} yield not only the perturbative 
$\beta$-function for a 
non-Abelian gauge theory coupled to fermions for arbitrary gauge fixing 
parameter $\xi$, but results beyond the 1-loop level as well.

\section{Gauge invariant flows}\label{gaugeinv}
Let us now come back to the question about gauge invariant flows. The standard 
formulation, based on the flow and mWI necessitates the introduction of some
gauge non-invariant operators for non-vanishing $k$, like a gluonic mass term,
in order to assure gauge invariant physical Green function. It would be 
interesting to see under which circumstances gauge non-invariant operators 
can be avoided.

The important observation to that end is, that usual gauge 
invariance for any scale $k$ 
can only be achieved by relaxing at least one of the key assumptions 
leading to the flow equation itself. 
There are essentially three options available. 
One either relaxes the constraints regarding the regulator function 
$R_k$ \cite{simionato}, or starts with a completely different mechanism 
for introducing the regularisation in the first place \cite{morris}, 
or introduces auxiliary fields \cite{analytic2,reutwett,analytic1,u1}. 
We shall now discuss all these options in more detail.

\subsection{Momentum independent regulator}
Let us comment on the first option, that is to change the 
requirements regarding $R_k$. 
It is straightforward to observe that a necessary and 
sufficient condition 
for usual gauge invariance even during the flow is just the vanishing of 
the commutator $\big[R^\phi_k+ L^*{1\ov \xi} L,G_k^{\phi\phi^*}[\phi]\big]=0$ 
(see 
\eq{mwi}). The only solution 
to this constraint (apart from the necessity of a momentum independent 
gauge fixing) is $R^\phi_k=\propto k^{d-2 d_\phi}$ thus 
introducing mass terms proportional to the cut-off scale $k$. Even though 
this choice satisfies \eq{property1} (with $n=1$), 
the second condition \eq{property2}, 
which guarantees that the ultraviolet (UV) behaviour of the 
theory is unaltered in the presence of the cut-off term, 
is no longer satisfied. 

The kernel of the trace 
in the flow equation \eq{flow} is no longer peaked at momenta about $k$, 
if \eq{property2} is violated. 
Even more so, \eq{flow} is not well-defined as it stands and needs some 
additional 
UV renormalisation. To be consistent, this has to be done on the 
level of the effective action rather than on the level of the flow equation. 
Otherwise the connection between the flow equation and the original --even 
though only formal-- path integral becomes unclear. Furthermore, since this 
additional UV renormalisation has to be $k$-dependent, one may ultimately 
lose the 1-loop structure of the flow equation. This depends on how the 
actual renormalisation is done. 

Moreover the interpretation of the flow equation is now 
completely 
different from the original Wilsonian idea. The flow equation no longer 
describes a successive integrating-out of momentum modes, but rather a 
flow in the space of massive theories. Even though the suppression of low 
momentum modes still works at every step of the flow, all parameters of the 
theories change for {\it all} momenta larger than $k$. This can be considered 
as a loss of locality, in the sense mentioned earlier. 
It has also to be pointed out that the result is 
{\it not} what is usually denoted by a massive gauge theory. The difference 
stems from the fact that the cut-off term --in the Wilsonian approach-- is 
introduced {\it after} the gauge fixing has been done. In the case of 
a massive gauge theory the Fadeev-Popov mechanism is applied to the path 
integral, where the action already includes the mass term. 
The difference between these 
two approaches are those terms stemming from 
$\int \d g \exp -k^2 \Tr (A^g)^2$, where $g(x)$ is a space-time dependent 
gauge group element. These are exactly the terms which usually are made 
responsible 
for the breakdown of renormalisability in massive gauge theories. 
A simple way to see this is as follows: Introduce 
$\chi, g=e^\chi$ as a new field and do the usual power counting with respect 
to the fields $(A,\chi)$. 
Dropping these terms changes the content of the theory, and although it 
looks superficially like a massive gauge theory, it is {\it not}. 

Apart from these conceptual problems, it appears that 
a numerical implementation is 
essentially out of reach. The momentum integrals involved would 
receive contributions from {\it all} momenta larger than $k$ instead of only 
being peaked within a small momentum shell at about $p^2\sim k^2$. Note that 
the numerical applicability of the Wilsonian flow equation may be seen as 
 one of its most attractive features, and  losing it is a big loss for 
gaining formal gauge invariance during the flow. 

It should be mentioned that a mass-like regulator remains 
an  interesting option for a first approximative computation, 
consistency checks or 
conceptional issues, as it typically simplifies analytic calculations 
tremendously. For more 
involved and non-trivial truncations in general, one has to use more 
elaborate regulators, though. 

\subsection{Gauge invariant variables}
A more attractive possibility is the proposal to change the starting 
point of the derivation, but to stick to 
the 1-loop nature of the resulting flow equation. This can only be done 
by mapping the degrees of freedom from the original fields to another set 
in a non-linear way, e.g.\ to a representation in terms of Wilson 
loops \cite{morris}. It is worth noting that this particular procedure
\cite{morris} requires an additional UV renormalisation which has been done 
explicitly by 
introducing Pauli-Villars fields. It also requires the 
introduction of 
a second gauge field. All these auxiliary fields only decouple in the limit 
$k\to 0$. Moreover, at least one of these fields 
needs a mass-like cut-off which again spoils the locality of the flow 
as defined earlier. 

Thus the same pitfalls concerning the applicability as in the other approach 
do finally apply here as well. Nevertheless --even though for finite $k$ the 
number of field degrees of freedom is considerably enlarged-- 
it seems to be possible to track down the original
degrees of freedom at any cut-off scale $k$. 

\subsection{Background fields}
Finally, we will discuss the third option --which has actually been 
used first \cite{reutwett} -- and which consists in introducing auxiliary 
fields \cite{analytic2,reutwett,analytic1,u1}. This is nothing but the 
adaption of the well-known 
background field formalism to a theory where a cut-off term is present. 
The key point here is to introduce the covariant derivative with respect 
to the background field $\bar A$ wherever a plain derivative was used in 
$R_k$. As a consequence, the effective action now satisfies the usual 
Ward identity if {\it all} fields are gauge transformed (The particular gauge 
transformations needed differ slightly in the various approaches). 

Even though the flow equation is still peaked at (covariant) momenta 
about $k$ it is obvious that the dependence of the effective action on the 
background field is non-trivial. One has to employ an additional equation to 
track down this dependence, which adds up with the flow equation to a bigger 
set of non-trivial equations. This is how in this approach 
the necessity of applying a non-linear transformation in field space shows 
up. At least the usefulness for numerical 
implementations seems to be questionable.

In summary one may conclude that the price for gauge invariance during the 
flow is rather high for all these approaches, in particular 
in comparison to the standard approach using the mWI. 
When it comes to numerical applications the 
latter approach seems clearly preferable. 

\section{Quantum fields at non-vanishing temperature}
We will now discuss an application of the Wilsonian Renormalisation Group to 
thermal field theory. The aim is to show 
that, in contrast to the previous sections which dealt with {\it quantum} 
fluctuations, for {\it thermal} fluctuations a fully gauge invariant flow 
can be constructed.

\subsection{How does temperature enter a quantum field theory}
We shall start with some general remarks on quantum field theories coupled 
to a heat bath. The temperature is introduced via the compactification 
of the (imaginary) time direction. This implies that the fields $\phi$
 live in the space $[-1/T,1/T]\times \R^3$ rather than $\R^4$. In 
addition, this results in the imposition of periodic (anti-periodic) 
boundary conditions for the bosonic (fermionic) degrees of freedom. 

Having said that, it follows that the modes with momenta much larger 
than the temperature will not be aware of the altered boundary condition, 
i.e. they will behave like modes living again in $\R^4$, that is like 
zero temperature modes. Stated differently, the UV physics is unaltered, and 
in particular, no further UV divergences 
than those encountered for the bare action at vanishing temperature 
will be observed. The low momentum or soft modes, however, do feel 
the new boundary condition. It is precisely for this reason that 
temperature has to be considered as an IR phenomenon.

\subsection{How to detect thermal corrections}
In order to identify the effects imposed by temperature, we have to 
introduce a reference point. Typically, a physical observable computed 
for some fixed temperature has to be compared to that very same 
observable computed at some other temperature, say $T=0$. Only 
their difference can be given the desired physical meaning. Thus, 
the effects of temperature are detected as a difference between 
observables measured for different boundary (temperature) conditions. 
This amounts to the statement that the natural object to study is
\beq\label{diff}
\Ga_{T}[\phi]-\Ga_{T=0}[\phi] \ ,
\eeq
where $\Ga_T$ denotes the 
effective action of a given quantum field theory at some fixed 
temperature $T$. Note that the proper definition of \eq{diff} requires an  
appropriate definition of the space of fields $\phi$. 
In other words, the finite temperature fields $\phi$ for which 
\eq{diff} makes sense 
have to be properly embedded in the zero temperature space-time 
\cite{finiteT}. Also physical observables derived from $\Ga_T$ may be of 
interest. A very important example is the thermal pressure
\beq
P[T]-P[0]\ .
\eeq 
It is related to the minimum of the effective potential, which 
itself is the leading order term within a derivative expansion 
of $\Ga_T$. For theories with their potential minimum at vanishing 
field, the pressure corresponds precisely to the field independent 
part of \eq{diff}.

\subsection{How to compute thermal corrections}
The more difficult question is now how \eq{diff} actually can be 
computed. Consider for example the thermal pressure: A perturbative 
computation of the pressure for a scalar field theory faces strong 
IR divergences. A systematic resummation has to be performed in order 
to obtain a finite result. Even worse is the case of QCD. Its perturbative 
computation is severely limited due to the non-perturbative magnetic 
sector of QCD. 
 These are generic examples for the problems of 
perturbative loop expansions at finite temperature, which do typically 
encounter serious IR problems at some loop order. 
In addition, 
it also has been observed that the convergence of the series for the 
thermal pressure is rather poor even in the domain, where perturbation 
theory should be applicable.

\section{Wilsonian flow for the thermal fluctuations}

We want to construct, based on Wilsonian ideas,
 a non-perturbative resummation procedure to compute 
\eq{diff}. Applications of Wilsonian flows to thermal field theories are 
not new \cite{TFT}. The usual approach is as follows: One constructs a 
flow equation for $\Ga_{k,T}$ which has to be solved (in some 
approximation) to yield $\Ga_T$. Within imaginary time, the flow 
equation is obtained from \eq{flow}, replacing the trace by
\beq\label{tr}
\Tr= \sum_{\phi}\int\0{d^3p}{(2\pi)^3}T\sum_n
\eeq
i.e. a sum over all momenta, Matsubara modes, and all (bosonic) 
fields and their indices. The substitution $p_0\to 2n\pi T$ for 
bosons is also understood. Then, one uses either the same flow but 
with $T=0$, or some alternative method like resummed perturbation 
theory, to compute $\Ga_{T=0}$. Combining these results, one finally 
obtains \eq{diff}. 

\subsection{Flow equation for the thermal contribution}
In contrast to the standard approach, we now aim at a flow equation not 
for $\Ga_{k,T}$, but directly for 
\beq\label{diffk}
\De\Ga_{k,T}[\phi]=\Ga_{k,T}[\phi]-\Ga_{k,0}[\phi]\ .
\eeq
The difference  \eq{diffk} effectively projects-out the thermal fluctuations. 
In the IR limit $k\to 0$, \eq{diffk} reduces to \eq{diff}, which is precisely 
what we are looking for. Note that in order to evaluate \eq{diffk}, we have 
fixed the temperature $T$ for all fields occurring in \eq{diffk}. 
Given the flow \eq{flow}, it is straightforward to write down a corresponding 
flow for \eq{diffk}, which reads for bosonic fields
\bea
\partial_t\De\Ga_{k,T}[\phi]= \partial_t \De\ln\CN_{k,T} +\012
\sum_\phi \int\0{d^3p}{(2\pi)^3}
&&\Bigl\{T\ \sum_{n} G^{\phi\phi^*}_{k,T}[\phi] \partial_t 
R^\phi_k\nonumber \\ \label{flowdiff}
&&-\int \0{d p_0}{2\pi} 
\, G^{\phi\phi^*}_{k,0}[\phi] \partial_t R^\phi_k\Bigr\}
\ .\eea 
It is worth pointing out the r$\hat{\rm o}$le 
of the normalisation constant $\CN_{k,T}$ 
for \eq{diffk}. In the standard approach \eq{flow}, it can simply be 
neglected because it corresponds to a shift of the zero point 
energy. However we have already encountered an example at zero temperature 
where it is essential 
to take into account the flow of the normalisation 
$\CN_k$ (see section~\ref{appl}). There it was related to the particular 
choice of the regulator $R_k$. At finite temperature, however, $\De\Ga_{k,T}$ 
at vanishing field measures a zero point energy {\it difference}, 
thus a physical observable. This implies that the 
normalisation properly has to be taken into account. 

However, up to now little has been gained while studying \eq{flowdiff} 
instead of \eq{flow}. For a general regulator $R_k$, the same 
qualitative problems regarding gauge invariance for arbitrary scale $k$ 
are encountered as at vanishing temperature. We shall now argue, that 
studying \eq{flowdiff} allows us to relax slightly one of the condition 
on the regulator function.

\subsection{Momentum independent regulator}
It was discussed earlier that a mass-like (i.e. momentum independent) 
regulator, employed for \eq{flow}, is not viable, the reason being the 
uncontrolled large momentum contribution to the flow. In \eq{flowdiff}, 
the situation is now different. Consider a mass-like regulator $R_k^\phi=k^2$ 
(and a momentum independent gauge fixing), for which the flow is given by
\cite{finiteT}
\beq\label{massdiff}
\0{\partial\De\Ga_{k,T}[\phi]}{\partial {k^2}}= \012
\sum_\phi \int\0{d^3p}{(2\pi)^3}
\left\{T\ \sum_{n} G^{\phi\phi^*}_{k,T}[\phi]-\int \0{d p_0}{2\pi} 
\,G^{\phi\phi^*}_{k,0}[\phi]\right\}+
\partial_{k^2}\De\ln\CN_{k,T} \ . 
\eeq 
First of all, the flow \eq{massdiff} 
is well defined in the IR limit. This is so, because using a mass-like 
regulator indeed cures the IR behaviour for the individual flows in 
\eq{massdiff}, and computing the difference does not change this 
property. The important observation is that the flow is also well-defined 
in the UV limit. This is so because the remaining UV divergences, which 
are not eliminated by the mass-term regulator, are eliminated through 
the subtraction of the $T=0$ counter part. Note that 
this implies that also for the initial scale $k_0$ the difference 
$\Delta_{k,T}\Gamma$ has to be local in the sense as defined in 
section~\ref{local}. Thus, for large momenta, the r.h.s. of 
\eq{massdiff} does not feel the presence of a thermal bath. Stated
differently, in the 
standard approach, the decay of $\partial_t R_k$ for large momenta 
ensures the UV finiteness of the flow equation. For a mass-like regulator 
$\partial_t R_k$ is proportional to $R_k$ and in particular not 
suppressed for large momenta. However, the suppression now comes 
from the cancellation at high momentum between the two propagators.  
This establishes that \eq{massdiff} is a well-defined {\it Wilsonian} 
flow for thermal fluctuations. 

It is worth mentioning that the cancellation of UV divergences is quite 
similar to the one employed in the BPHZ-procedure. The subtraction of possibly 
divergent terms takes place on the level the integrand rather than on the 
level of the regularised full expressions. This is how the explicit 
introduction of an UV renormalisation scale is avoided. 

\subsection{Gauge invariance}
Let us now see what happens in the case of pure gauge theories. 
We consider a SU($N$) gauge theory in an axial gauge $(U_\mu A_\mu)$. 
As mentioned before 
it is also necessary to take a momentum independent gauge 
fixing parameter $\xi$, which is 
known to have  at $\xi =0$ a non-perturbative fixed point \cite{Axial}. 
The mWI  
for $\Ga_{k,T}[A]$ is derived from \eq{mwi} for a general regulator as  
\beq 
\del_\alpha \Gamma_{k,T}[A]=\di \!\! 
\int\0{d^3p}{(2\pi)^3}T\sum_n
\left(\frac{1}{U^2\xi} U_\mu\partial_\mu \alpha\ U_\nu A_\nu  -\0g2\,
[\alpha, R_k^{A}] 
G^{AA}_{k,T}[A]\right). 
\label{mWI-T} 
\eeq
The mWI for \eq{flowdiff} follows straightforwardly from \eq{mWI-T} as
\bm{c} 
\del_\alpha\De\Gamma_{k,T}[A]
= \0g2 
\!\int\! \0{d^3 p}{(2\pi)^3}\left(\!\int\0{dp_0}{2\pi} [\alpha, R_k^{A}] 
G^{AA}_{k,0}[A]-T\sum_n [\alpha, R_k^{A}] 
G^{AA}_{k,T}[A] \right).\ \ \ {}
\label{mWIdiff-T} 
\eg
Note, that the term in \eq{mWI-T} generated from the gauge fixing has 
canceled, and the only remaining breaking of gauge symmetry comes from the 
regulator function. Within the derivation of (22) it is important to use the 
correct boundary conditions for $\alpha$ (for details see \cite{finiteT}). 
However, with the mass-like regulator as used in \eq{massdiff}, the mWI reads
\beq 
\del_\alpha\De\Gamma_{k,T}[A]\label{mWImassdiff-T} =0,
\eeq
because the commutator $[\alpha, R_k^{A}]$ vanishes for momentum independent 
regulator. Note, that the WI controls only 
the gauge invariance for the field dependent part of the 
effective action. As mentioned before the normalisation has to be properly 
chosen which is a well-known problem in gauge theories at finite temperature 
\cite{bernard}. As a result $\De\ln\CN_k[\xi]$ is 
$\xi$-dependent. It turns out 
that it is precisely this $\xi$-dependence which guarantees 
that the field independent part of \eq{flowdiff} is 
$\xi$-independent and thus gauge independent. This of course
 ensures gauge independence for the 
field independent part
$\De\Ga_{k,T}[0]$. 

We have established the following important result: The flow 
equation \eq{massdiff} is a) a well-defined flow in the Wilsonian 
sense and b) explicitly maintains gauge invariance for arbitrary scale $k$.

\subsection{Remarks}
We will finish this section with some remarks. 
Note first that studying \eq{diff} makes an
additional implicit assumption, which is that the physical degrees of freedom 
describing
the zero temperature theory are assumed to remain reasonable 
degrees of freedom at 
temperature $T$. Of course, this is the prerequisite to give the 
difference \eq{diffk} a meaning. This does not exclude the introduction 
of composite operators at some intermediate scale as well. In the present
formulation,  this 
has  to be done both for the $T=0$ and the $T\neq 0$ theory at some fixed 
scale $k_b$.  

A second comment concerns the initial conditions.
In contrast to the usual approach, the flow for \eq{diffk} can no longer 
be a functional differential equation for $\De\Ga_{k,T}$ only. This is 
evident from the definition of $\De\Ga_{k,T}$ in the first place. The 
r.h.s. will necessarily imply information about the underlying theory, 
for which the thermal fluctuations are computed. This is reflected by 
the fact that the flow equation will depend as well on $\Ga_{k,0}$. 
Alternatively, this can be deduced from the initial condition. In 
the UV limit, we have
\beq \label{UV}
\lim_{k\to\infty}\De\Ga_{k,T}[\phi]=0
\eeq
while the IR limit reads
\beq
\lim_{k\to 0}\De\Ga_{k,T}[\phi]=\Ga_T-\Ga_{T=0}\ .
\eeq
\Eq{UV} implies, that, in order to specify the theory, one has to 
furnish some information regarding the theory at vanishing temperature, 
that is on $\Ga_{T=0}$. 
This is very similar to another, but qualitatively different proposal 
within real-time thermal field theory by d'Attanasio and Pietroni \cite{TRG}. 
There, the starting point needs the renormalised $T=0$ theory as 
an input as well.

Finally, it is worth noting the similarity between our flow \eq{massdiff}, 
and a recent proposal to compute the thermal pressure for scalar 
theories \cite{Pressure}. There, the authors presented an IR finite 
resummation formula for the thermal pressure, which is expressed in 
terms of a mass integral. If we restrict ourselves to scalar fields 
only, and to the leading order in a derivative expansion, then our 
flow \eq{massdiff} corresponds precisely to their proposal. In this 
sense, our flow is the generalisation to gauge theories and the 
entire effective action. 

\section{Discussion and outlook}

For the integrating-out of {\it quantum} fluctuations, the present status 
is as follows: There are several possibilities to rescue usual 
gauge invariance even during the flow, but all of them have to face 
serious problems -- at least when (numerical) applicability is demanded. The 
reason is that the only way to gain gauge invariance within a Wilsonian 
formulation is to employ non-linear transformations in field space or even 
to lose locality --as defined in section~\ref{gaugeinv}-- of the 
flow equation. The resulting formulations are quite difficult to handle 
if it comes to non-trivial applications. The technical price for  
gauge invariance along the flow is quite big as opposed to the approach
using the mWI. The handling of the latter, in particular, does not need 
other techniques than those already used for the flow equation itself. 

It is therefore feasible to employ --for more involved questions-- 
the regulators $R_k$ satisfying the conditions 
\eq{property1},\eq{property2} along with the mWI to guarantee gauge 
invariance for physical Green functions. After all, the flow equation should
rather be seen as a well-defined procedure as to how to calculate the full 
quantum effective action. This does not necessitate usual gauge invariance 
{\it during} the flow but the {\it control} of gauge invariance for the 
physical Green functions, that is, at $k=0$.  

All the approaches mentioned in section~\ref{gaugeinv} certainly 
have their merits for particular applications and conceptional issues. Further 
investigations in these directions are needed in order to clarify 
some outstanding conceptional questions, e.g. concerning perturbative 
$\beta$-functions beyond 1-loop \cite{analytic2}, 
the full-fledged 
inclusion of topological non-trivial configurations \cite{u1} 
and anomalies \cite{bonvian}. 

For the integrating-out of {\it thermal} fluctuations, we have been able 
to construct a gauge invariant renormalisation group equation, based on 
a mass-like regulator term in combination with an axial gauge fixing. 
The proposal is very similar to --and can be seen as a generalisation to gauge 
theories of-- an approach advocated earlier for scalar 
theories\cite{Pressure}. This proposal differs qualitatively from 
another gauge-invariant thermal 
renormalisation group introduced earlier \cite{TRG}. 

The mass-like regulator typically allows analytic computations, as has been 
done for the thermal pressure for scalar theories \cite{Pressure}. 
Further applications in thermal 
field theory should include the computation of the thermal pressure for a 
gas of  bosons and gluons. In a first approximation, it suffices to insert 
the classical action $S$ at $T=0$ into the flow 
equation instead of $\Ga_{k,T=0}$. The higher order corrections are then 
obtained through the 
back coupling between an improved action at vanishing temperature, 
and the flow equation.

More generally, it would be very interesting to see whether the Hard 
Thermal Loop effective action could be used as a starting point, rather 
than the $T=0$ one. This might be a good starting point to go beyond the 
HTL approximation. We hope to report on these matters in the future.
  
\section*{Acknowledgments}
It is a pleasure to thank A.~Andrianov, B.~Dolan, U.~Ellwanger, F.~Freire, 
J.I.~Latorre and C.~Wetterich for discussions, and the organisers for a 
stimulating conference and financial support.


\section*{References}
\begin{thebibliography}{99}
\def\BOOK#1#2#3#4{#1 {\sc #2}, #3, #4}
\def\PRA#1#2#3#4#5{ #1   Phys.~Rev.~{\bf A #3} (19#4) #5}
\def\PRB#1#2#3#4#5{#1   Phys. Rev.~{\bf B #3} (19#4) #5}
\def\PRL#1#2#3#4#5{#1   Phys. Rev.~Lett.~{\bf #3} (19#4) #5}
\def\PRC#1#2#3#4#5{#1   Phys. Rev.~{\bf C #3}  (19#4) #5}
\def\PRD#1#2#3#4#5{#1   Phys. Rev.~{\bf D #3} (19#4) #5}
\def\PRE#1#2#3#4#5{#1   Phys. Rev.~{\bf E #3} (19#4) #5}
\def\PRep#1#2#3#4#5{#1   Phys. Rep.~{\bf  #3} (19#4) #5}
\def\NPB#1#2#3#4#5{#1   Nucl. Phys.~{\bf B #3} (19#4) #5}
\def\PLB#1#2#3#4#5{#1   Phys. Lett.~{\bf B #3} (19#4) #5}
\def\ibid#1#2#3#4#5{#1   {\it ibid.~}{\bf #3} (19#4) #5}
\def\PTP#1#2#3#4#5{#1   Prog. Theor.~Phys.~{\bf B #3} (19#4) #5}
\def\SSC#1#2#3#4#5{#1   Solid State Comm.~{\bf  #3} (19#4) #5}
\def\EPL#1#2#3#4#5{#1   Europhys. Lett.~{\bf #3} (19#4) #5}
\def\JCP#1#2#3#4#5{#1   J.~Phys. (Paris) {\bf  #3} (19#4) #5}
\def\JPA#1#2#3#4#5{#1   J.~Phys. {\bf A  #3} (19#4) #5}
\def\JPB#1#2#3#4#5{#1   J.~Phys. {\bf B  #3} (19#4) #5}
\def\JPC#1#2#3#4#5{#1   J.~Phys. {\bf C  #3} (19#4) #5}
\def\ZPC#1#2#3#4#5{#1   Z.~Phys. {\bf C  #3} (19#4) #5}
\def\JETP#1#2#3#4#5{#1   Soviet Physics JETP Lett.~{\bf #3} (19#4) #5}
\def\MPLA#1#2#3#4#5{#1   Mod.~Phys. Lett.~{\bf A  #3} (19#4) #5}
\def\PA#1#2#3#4#5{#1   Physica {\bf A  #3} (19#4) #5}
\def\PS#1#2#3#4#5{#1   Physics {\bf   #3} (19#4) #5}
\def\AP#1#2#3#4#5{#1   Ann. Phys. {\bf  #3} (19#4) #5}
\def\IJMPA#1#2#3#4#5{#1   Int.~J. Mod. Phys.~ {\bf A  #3} (19#4) #5}
\def\LNC#1#2#3#4#5{#1   Lett.~Nuevo Cimento {\bf   #3} (19#4) #5}
\def\PPR#1#2{#1  {\tt #2}}
\def\and#1#2#3{{\bf #1} (19#2) #3}
\bibitem{Wilson}K.G.~Wilson and I.G.~Kogut, Phys. Rep. {\bf 12} (1974) 75;
F.~Wegner and A.~Houghton, Phys. Rev. {\bf A 8} (1973) 401.
\bibitem{ERG} J.~Polchinski, Nucl. Phys. {\bf B 231} (1984) 269.
\bibitem{heavy}U.~Ellwanger, M.~Hirsch and A.~Weber, Z. Phys. {\bf C 69}
 (1996) 687; Eur. Phys. J. {\bf C1} (1998) 563. 
\bibitem{Axial}\PLB{D.F.~Litim and J.M.~Pawlowski,}{}{435}{98}{181} 
[{\tt hep-th/9802064}]; {\tt hep-th/9809023}.
\bibitem{analytic2}D.F.~Litim and J.M.~Pawlowski, under completion.
\bibitem{reutwett}\NPB{M.~Reuter and C.~Wetterich,}{}{391}{93}{147}; 
\and{B 408}{93}{91}; \and {B 417}{94}{181}; 
F.~Freire and C.~Wetterich, Phys. Lett. {\bf B 380} (1996) 337.
\bibitem{collect}M.~Bonini, M.~D'Attanasio and 
G.~Marchesini, {Nucl.~Phys.}~{\bf B 421} (1994) 429; 
U.~Ellwanger, Phys. Lett. {\bf B 335} 
(1994) 364. 
\bibitem{analytic1}D.F.~Litim and J.M.~Pawlowski, {\tt hep-th/9809020}. 
\bibitem{datmorris}\PLB{M. d'Attanasio and T. Morris,}{}{378}{96}{213}. 
\bibitem{berwett}B.~Bergerhoff and C.~Wetterich, Phys. Rev. {\bf D 57} 
(1998), 1591. 
\bibitem{simionato}M.~Simionato, {\tt hep-th/9809004}, {\tt hep-th/9810117}. 
\bibitem{morris}T.~Morris, these proceedings [{\tt hep-th/9810104}]. 
\bibitem{u1}J.M.~Pawlowski, Phys. Rev. {\bf D 58}:045011 (1998).
\bibitem{TFT}D.F.~Litim, {\it Wilsonian flow equations and thermal field 
theory}, {\tt hep-ph/9811272}, and references therein. 
\bibitem{TRG}\NPB{M.~D'Attanasio and M.~Pietroni,}{}{472}{96}{711}, 
\NPB{}{}{498}{97}{443}.
\bibitem{Pressure}\PLB{I.T.~Drummond, R.R.~Horgan, P.V.~Landshoff and 
A.~Rebhan,}{}{398}{97}{326}; \NPB{}{}{524}{98}{579}; \PPR{D. B\"odeker, 
P.V.~Landshoff, O.~Nachtmann, and A.~Rebhan,}{hep-ph/9806514}; \PPR{A.~Rebhan,}
{hep-ph/9808480}.
\bibitem{finiteT}D.F.~Litim and J.M.~Pawlowski, under completion.
\bibitem{bernard}\PRD{C.W.~Bernard,}{}{9}{74}{3312}. 
\bibitem{bonvian}M.~Bonini, F.~Vian, Nucl. Phys.{\bf B 511} (1998) 479; Nucl. 
Phys. {\bf B 532} (1998) 473; F.~Vian, these proceedings 
[{\tt hep-th/9811055}].

\end{thebibliography}
\end{document}